\documentclass[12pt]{article}
\newtheorem{thm}{Theorem}

\begin{document}

 \begin{center}

{\Large  The Levels of Quasiperiodic Functions on the Plane,
Hamiltonian Systems and Topology}

\end{center}

\vspace{0.3cm}

\centerline{\bf S.P. Novikov \footnote {\it University of Maryland,
IPST and Math Department, College Park, MD, 20742-2431 and Moscow
117940, Kosygina 2, Landau Institute for Theoretical Physics, e-mail
novikov@ipst.umd.edu; This work is going to appear in the Russian Math Surveys
 (Uspekhi Math Nauk), 1999, n 6. It is partially supported by the NSF Grant
DMS 9704613}}

\vspace{0.1cm}

{\bf Abstract:}
{\it Topology of levels of the quasiperiodic functions with $m=n+2$ periods
on the plane is studied.
For the case of  functions with $m=4$ periods   full description is obtained
for the open everywhere dense family of functions.
 This problem is equivalent
to the study of Hamiltonian systems on the $n+2$-torus
 with constant rank 2 Poisson
bracket. In the cases under investigation we proved that this system is topologically
completely integrable in some natural sence where interesting integer-valued
locally stable topological characteristics appear. The case of 3 periods has been extensively
studied last years by the present author, Zorich, Dynnikov and Maltsev
for the needs of solid state physics (''Galvanomagnetic Phenomena
in Normal Metals'');
The case of 4 periods might be useful for the Quasicrystals.}

\vspace{0.1cm}

Let us consider a periodic function $f(x), x=(x^1,\ldots,x^m),n=m+2,$
 in the space $R^{n+2}$ with the group of periods $Z^m\subset R^m,$ i.e.
  the function on the torus
 $f:T^m\rightarrow R, T^m=R^m/Z^m$. We may think  that the lattice
 $Z^m$ is generated by the standard basic vectors
  $e_j=(0,\ldots,0,1,0,\ldots)$. For every plane $R^2\subset R^{n+2}$
  with linear coordinates $(y^1,y^2)$
  given by the system of the linear equations $l_i=b_i,l_i\in R^{m*},
  i=1,\ldots,n,$ we have a restriction $g(y)$ of the function $f(x)$
  on the plane $g_l(y)=f(x(y))$. Such a function $g_l(y)$ is called
  {\bf Quasiperiodic} with $m$ periods on the plane.

  {\bf Problem:} What can be said about Topology of the levels $g_l(y)=const$?

  The case $m=3$ has been extensively studied in connection with
  the Galvanomagnetic Phenomena in the single crystal normal metals
  since late 50s by the school of I.Lifshitz (M.Azbel, M.Kaganov, V.Peschanski
  and others-see the survey and references in  \cite{NM}).
  Topological investigations
  of this problem have been started after the  article \cite{N}
  in 1982 in the present authors seminar
  (see \cite{NM,N,Z,D,N2} for the main topological and physical results
   of our group).

  The space $T^3=R^3/Z^3$ here is a space of quasimomenta, $Z^3$ is a
  reciprocal lattice, the function $f(x)$ is a dispersion relation.
  The standard notation for quasimomenta would be $k$ or $p$ but we
  use the letter $x$ for them here. The most important is  Fermi level
  $f=\epsilon_F$ which is a {\bf Fermi surface} in $R^3/Z^3$.
   In the magnetic field $B$ ''semiclassical''
  electrons move in the planes $R^2$ orthogonal to $B$ in the space of quasimomenta.
    This family of planes
  leads to the quasiperiodic functions with 3 periods in the planes.

  The case $m=4$ is the main subject of this work. Our results may be useful for
  the theory of quasicrystals. By definition, the {\bf Rational Plane} is one
  given by two linear equations $l_1=b_1, l_2=b_2$ where both linear forms
  $l_1,l_2$ have rational coefficients in the standard basis of lattice above
  (no restrictions for right-hand
  parts $b_1,b_2$).

  \begin{thm} There exist nonempty open neighborhoods $U_j\subset RP_3$ of the
  rational directions $l_j\in U_j, j=1,2$ such that:

  For every  directions $l'_j \in U_j,j=1,2$ the connectivity component of
  the level $g_{l'}(y)=const$ is either compact  or
  belongs to the strip of finite width between two parallel lines in $R^2$.
  Except codimension one subset of the ''nongeneric levels''
  this situation is stable under the variation of
   all parameters involved (including variation of the directions $l'$
   and of
   the periodic function $f$ in $R^4$). The direction of the strip is an
   intersection of the 2-plane $l'$ with some 3-dimensional hyperplane
   $R^3\subset R^4$ which is integral in the standard lattice basis above
   and stable under the small variations of parameters.
   \end{thm}

We may consider this problem from the different point of view.
Let Poisson Bracket is given on the torus $T^4$
 which is {\bf constant and degenerate}
with Annihilator (Casimir) generated by 2 multivalued fuctions $l'_1,l'_2$.
 {\bf The Hamiltonian function} $f(x)$ generates a flow whose
trajectories are equal to our curves---sections of the level $M^3_a:f=a$
by the family $l'$ of 2-planes $l'_1=b_1,l'_2=b_2$.
 Remove from the level $M^3_a$
all compact nonsingular trajectories (CNST):
$$M^3_a=(CNST)\bigcup_iM_i$$

Now fill in all boundaries of $M_i$ by the family of
2-discs in the corresponding 2-planes whose boundaries
 are the separatrix trajectories. We are coming to the piecewise smooth
 3-manifolds $\bar{M}^3_i\subset T^4$ representing the 3-cycles $z_i\in
 H_3(T^4,Z)$.
 Under the same restrictions as in the previous theorem, we have following

 \begin{thm}
All cycles $z_i$ are nonzero in $H_3(T^4,Z)$, equal to each other
 up to the sign, and sum
of them is equal to zero. Every manifold $\bar{M}^3_i$ and
corresponding cycle $z_i$
is represented as
an image of the 3-torus $T^3\rightarrow T^4$. These cycles are stable under the small
variation of parameters.

\end{thm}

The idea of the proof is following. Take any pair of rational directions
$l_1,l_2$. Assume that $l_1=x^4$, and the corresponding levels are tori
$x^4=c$. Take small enough generic variation $l'_2$ of the direction $l_2$.
The intersections of hyperplanes $l'_2=b_2$ with tori $T^3_c:x^4=c$
leads us to the one-parametric family of problems previously solved \cite{Z}
about the plane sections of the Fermi surfaces $M^3_a\bigcap T^3_c$
in the 3-tori $T^3_c$. We may meet also a finite number of the
singular sections $M^3_c$. In the nonsingular sections $M^3_c$
there are compact trajectories  (i.e. closed
in the universal covering space $R^3$) and  open trajectories lying
on the 2-tori $T^2_{i,c}\subset T^3_c$ nonhomologous to zero in the homology
group $H_2(T^4,Z)$.
All other trajectories are the singular limits of these types.

Both these types are topologically stable. So we may
have one-parametric family of compact trajectories or one-parametric
family of 2-tori. Let this family be generic.
  The nonsingular compact trajectories remain
 compact after small perturbations (in particular, after replacing the direction $l_1$
 by $l'_1$). They will be removed from our manifold according to the construction
 above.

 One-parametric family of 2-tori $T^2_{i,c}$ may have generic singularities.
 The pair of 2-tori may meet each other in the nonsingular level $c$.
 The structure of this process can be extracted from the arguments given in
 the work \cite{D}: it is the same picture as one when the pair of 2-tori
 are meeting each other in the generic boundary point of the stability zone.
 These tori have the opposite homology classes, so they are homotopic in
 the torus $T^3$. After meeting they leave a tail consisting of the compact
 orbits. So all our picture is covered by the map $T^2\times I\rightarrow
  M^3_c$. It looks like 2-torus ''turned back'' being reflected by this level.
 Another situation might happen when 2-torus meets a singular level $M^3_c$
 (i.e. the closed 1-form $dx_4$ has a Morse critical point). The local
 and minima play no role here. We concentrate on the case of the Morse
  index equal to 2 (the case of index 1 corresponds to the inverse process).
   The vanishing
  cycle on the torus $T^2_c$ should be homologous to zero in $T^2$ because
  it is homologous to zero in $T^4$. By that reason we shall come to
  the union $T^2\bigcup S^2$ after surgery. However all plane sections
  of the topological 2-sphere are compact. So our torus passed
  critical level homotopically
   unchanged giving raise to the tail of compact orbits. Let us mention that
   such tails of compact orbits might appear also from
   the more trivial reasons.
   Now we perturb the direction $l_1$ replacing it by the generic
small   perturbation $l'_1$. Cutting off all compact nonsingular
sections and filling in all boundary separatrix compact curves by 2-discs
in the corresponding plane,we are coming to the natural images of the
3-tori $T^3\rightarrow \bar{M}^3\subset T^4$.
 Here the map is monomorphic in homology groups. After that  both our
 theorems follow from the same arguments as before, for the 3-dimensional case.

{\bf Remarks}: 1.It  became clear after discussion of the present author with
Dynnikov
that {\bf no generalization of this results
is possible for the number of periods more than 4 for the
 directions not closed
to the rational ones}.
 2.For the case of 4 periods our {\bf conjecture} is that
Theorems 1 and 2 are valid for the measure one family of 2-planes in the
 grassmanian, but in generic case the Hausdorf codimension of the
  exceptional set
is less than one. For the even (and therefore nongeneric)
cases like $\sum_i\cos(x^i)=0$ 
these theorems are probably not true.
 3.Some generalization of our results for the directions
closed to
the rational one is probably possible for any number of periods.

\end{document}